\documentstyle[epsf,eqsecnum,floats,preprint,aps]{revtex}

\tighten

\begin{document}
 
\newcommand{\vp}{\varphi}
\newcommand{\f}{{\rm out}}
\newcommand{\la}{\langle}
\newcommand{\ra}{\rangle}
\newcommand{\WW}{{\rm\bf{WW}}}
\newcommand{\WWW}{{\rm\bf{WWW}}}
\newcommand{\al}{\alpha}
\newcommand{\be}{\begin{equation}}
\newcommand{\ee}{\end{equation}}
\newcommand{\bea}{\begin{eqnarray}}
\newcommand{\eea}{\end{eqnarray}}
\newcommand{\PSbox}[3]{\mbox{\rule{0in}{#3}\includegraphics{#1}\hspace{#2}}}

\title{CP Violation from Surface Terms in\
the Electroweak Theory without Fermions}

\author{Arthur Lue$^{1}$\footnote[1]{\tt lue@cuphyb.phys.columbia.edu} and 
Mark Trodden$^{2}$\footnote[2]{\tt trodden@theory1.phys.cwru.edu.}}

\address{~\\$^{1}$Department of Physics \\
Columbia University \\
New York, NY 10027, USA.}
\address{~\\$^{2}$Particle Astrophysics Theory Group \\
Department of Physics \\
Case Western Reserve University \\
10900 Euclid Avenue \\
Cleveland, OH 44106-7079, USA.}

\maketitle

\begin{abstract}

We consider the effect of adding a CP-odd, $\theta F\tilde{F}$-term to
the electroweak Lagrangian without fermions. This term
affects neither the classical nor perturbatively quantum physics, but 
can be observed through non-perturbative quantum processes. We
give an example of such a process by modifying the theory so that it
supports Higgs-winding solitons, and showing that the rates of decay of
these solitons to specific final states are CP violating.  We also discuss
how the CP symmetry is restored when fermions are included.

\end{abstract}

\setcounter{page}{0}
\thispagestyle{empty}

\vfill

\noindent CU-TP-879	\hfill 

\noindent CWRU-P9-98 \hfill

\noindent hep-ph/9802281 \hfill Typeset in REV\TeX

\eject

\vfill

\eject

\baselineskip 24pt plus 2pt minus 2pt

\section{Introduction}

In Quantum Chromodynamics (QCD) it is well known that violation of the
charge-parity (CP) symmetry can appear through the inclusion of a
CP-odd total divergence

\be \Delta S_{QCD} = \frac{\theta_{QCD}}{16\pi^2} \int d^4x\, {\rm
Tr}\left(F_{\mu\nu}{\tilde F}^{\mu\nu}\right)
\label{QCD}
\ee 
in the action\cite{{tHooft},{JR}}. Here $F$ is the $SU(3)$ field
strength tensor and $\theta_{QCD}$ is a dimensionless parameter. This
modification does not affect the classical equations of motion and
introduces no additional Feynman graphs in perturbation theory.
It is only through nonperturbative quantum phenomena that this CP-odd
surface term can be observed. In particular, the electric dipole
moment of the neutron is affected by the presence of this term.
Nevertheless, no dipole moment is observed and the tight experimental
upper bound on this quantity translates into the constraint
$\theta_{QCD} \leq 10^{-9}$\cite{dipole}
.

The manifestation of~(\ref{QCD}) in physical observables is
complicated.  Calculations deriving the relevant effects invoke
dynamical chiral symmetry breaking and other phenomenological insights
into strong physics.  It seems useful to identify other manifestations
of this type of total divergence to ascertain how they work.

If one introduces a term analogous to~(\ref{QCD}) in the standard
electroweak theory, with $F$ being the $SU(2)$ gauge field strength,
it has no observable effect.  The chiral coupling of fermions to the
SU(2)-gauge fields allows the term to be eliminated by performing a
phase rotation on quark and lepton fields\cite{Fujikawa}.  Thus a
total divergence of the form (\ref{QCD}) is unobservable in the
standard electroweak theory.

Nevertheless, if one considers the electroweak theory without
including fermions, then it should be possible to observe the total
divergence. We will look for CP violating effects which arise from
this term.  Generally it is difficult to demonstrate such effects
unless nonperturbative phenomena are identified. In this paper we
alter the electroweak theory so that it supports classically stable
solitons\cite{Skyrme,GipTze,A&R 85}, thereby introducing
nonperturbative objects into the spectrum.  We show that the rate of
decay of these solitons to specific final states is CP violating and
briefly discuss how the CP symmetry is restored when fermions are included.

\section{Preliminaries}
Consider the bosonic sector of the electroweak theory as an effective
field theory. In addition to the usual terms, assume that the Lagrangian 
density for this theory contains a gauge invariant, CP-even, 
Skyrme term \cite{Skyrme,GipTze} which stabilizes Higgs winding 
configurations as solitons. For simplicity consider only the 
$SU(2)$ gauge fields. The action is

\bea
	S_0[\Phi,A_\mu] = \int d^4x \left[
		-\frac{1}{2}{\rm Tr}\ F_{\mu\nu}F^{\mu\nu}
		+\frac{1}{2}{\rm Tr}\  D_\mu\Phi^\dagger D^\mu
		\Phi
		-\frac{\lambda}{4}\left({\rm Tr}\ \Phi^\dagger\Phi-v^2\right)^2
		+ \right. \nonumber	\\
	\left.\frac{1}{32e^2v^4} {\rm Tr}
	\left( D_\mu\Phi^\dagger D_\nu\Phi-
D_\nu\Phi^\dagger D_\mu\Phi\right)^2\right] \ ,
\label{S0}
\eea
where

\bea
	F_{\mu\nu} &=& \partial_\mu A_\nu - \partial_\nu A_\mu -
	ig[A_\mu,A_\nu]\ , \nonumber	\\
	 D_\mu\Phi &=&\partial_\mu\Phi - igA_\mu\Phi	\nonumber
\eea
and $\Phi(x)$ is related to the standard Higgs-doublet $(\vp_1, \vp_2)^T$ by
\be
	\Phi = \left(\matrix{\vp_2^* & \vp_1 \cr -\vp_1^* & \vp_2} \right).
\label{higgsmatrix}
\ee
Here $g=0.65$ is the gauge coupling constant, the Higgs vacuum
expectation value (VEV) is $v=247$GeV, and the gauge and Higgs boson
masses are $m =\frac{1}{2}gv$ and $m_H=\sqrt{2\lambda}v$ respectively.
Note that the action $S_0$ is invariant under the usual
CP-transformation.

Now consider adding the term

\be
\frac{\theta}{32\pi^2} \int d^4x\, \epsilon^{\mu\nu\alpha\beta}
{\rm Tr}(F_{\mu\nu}F_{\alpha\beta}) 
\label{ffdual}
\ee
to the action. In the absence of chirally coupled fermions, this term
cannot be rotated away, and we assume throughout this paper that there are no 
fermions present. In particular,~(\ref{ffdual}) is
CP-odd, and thus we expect there to be CP violating processes associated 
with the new term. Moreover, as a consistency check, such processes should 
no longer be present if one introduces fermions, and we shall 
demonstrate that this is the case.

\section{Soliton Decays}
Consider for the moment the action (\ref{S0}) in the limit where the
Higgs self-coupling $\lambda\rightarrow\infty$ and the gauge fields
decouple, $g\rightarrow 0$.  In this limit we recover the Skyrme
action \cite{Skyrme}.  Such a theory is known to support Higgs-winding
topological solitons

\be
	\Phi({\bf x}) = \frac{v}{\sqrt{2}}\ e^{-i\sigma^a\hat{x}^aF(r)},
\label{skyrmion}
\ee
where $F(r=0)=\pi$ and $F(r\rightarrow\infty)\rightarrow 0$.
Defining 

\be
U \equiv \frac{\Phi}{\sqrt{{\rm Tr}\Phi^\dagger\Phi/2}} \ , 
\ee
the winding
number is

\be
	w[\Phi] = \frac{1}{24\pi^2}\int d^3{\bf x}\
	\epsilon^{ijk}\left[(U^\dagger\partial_iU)(U^\dagger\partial_jU)
	(U^\dagger\partial_kU)\right],
\ee
and the configuration (\ref{skyrmion}) has $w[\Phi] = 1$.

Now, if we allow a finite Higgs self-coupling, one may identify a
sequence of configurations which connects the soliton to a classical
vacuum configuration.  All such paths must go through a configuration
where $\Phi=0$ at some point in space.  For this particular configuration, 
the winding number of the Higgs field is not defined.  If $\lambda$ is
very large, the soliton remains classically stable and any
configuration close to (\ref{skyrmion}), but for which $\Phi$ is not
on the vacuum manifold (i.e., where ${\rm Tr}\Phi^\dagger\Phi \ne
v^2$), has much larger energy than that of the soliton.  However, if $\lambda
\rightarrow 0$, the configuration (\ref{skyrmion}) is no longer stable,
and there exist sequences of configurations with monotonically
decreasing energy that connect (\ref{skyrmion}) to the vacuum.  These
opposite limits imply that there is a value $\lambda^*(e)$ of the
Higgs self-coupling for which the classical soliton is critically
stable.

If we restore the gauge fields, there exists a second class of
sequences of configurations that connect the soliton to a classical
vacuum configuration \cite{A&R 85,dH&F}.  For this class, the gauge
field aligns with the Higgs field such that they both have the same
winding.  In the limit where the gauge coupling is small, the soliton
(\ref{skyrmion}) is classically stable.  However, as the parameter
$\xi = \frac{4e^2}{g^2}$ decreases, the soliton becomes increasingly
heavy, in units of $m_W$.  Thus there exists a value for $\xi = \xi^*$ at
which the soliton becomes critically stable to gauge alignment.

Consider the classical configuration space of the action, where gauge 
equivalent configurations are identified.  We have identified two classes of
paths that connect the soliton to the vacuum:
\begin{enumerate}
\item Paths that go through a zero of the Higgs field.  Tunneling by
such paths is controlled by the Higgs mass,
$m_H = m_W\sqrt\frac{8\lambda}{g^2}$, since the
barrier height is controlled by $(\lambda-\lambda^*)$.

\item Paths that do not go through a zero of the Higgs field
(ie. paths
for which $\int F\tilde{F}\neq0$).  The barrier height in this direction
is controlled by the parameter $(\xi - \xi^*) > 0$.
\end{enumerate}

Now, let us identify the quantum
transition amplitude $\langle\f|{\cal T}|s\rangle$
between asymptotic states, where 
$|s\ra$ is the properly quantized soliton state and $|\f\rangle$ is a
state of a definite number of W particles built on top of a vacuum
configuration in unitary gauge. This amplitude has two components
\be
\langle\f|{\cal T}|s\rangle = \left(\int_{\hbox{class 1}}+
\int_{\hbox{class 2}}\right)[d\Phi][dA]
\exp\left(iS_0+\frac{i\theta}{16\pi^2} \int F\tilde{F}\right) \ .
\ee
The term $\frac{1}{16\pi^2}\int F\tilde{F}=0$ for paths of class 1, and
$\frac{1}{16\pi^2}\int F\tilde{F}=1$ for paths of class 2.
\footnote{Note, although $\int
F\tilde{F}$ is not well-defined \cite{5authors}, one can show that the approach
taken is equivalent to evaluating the transition amplitude in the
Hamiltonian formalism with a $\theta$-vacuum.}
The previous expression then becomes

\bea
\langle\f|{\cal T}|s\rangle & = & \int_{\hbox{class 1}}[d\Phi][dA]
e^{iS_0} +e^{i\theta}\int_{\hbox{class 2}}[d\Phi][dA]e^{iS_0} \nonumber \\
& \equiv & {\cal A}_1 + e^{i\theta}{\cal A}_2 \ ,
\label{basic}
\eea
where ${\cal A}_1$ and ${\cal A}_2$ are in general complex. Now compare this
process to the 
CP conjugate process. Since $S_0$ is CP invariant, we obtain
\be
\langle \overline{\f}|{\cal T}|{\bar s}\rangle = {\cal A}_1 + 
						e^{-i\theta}{\cal A}_2 \ ,
\label{antibasic}
\ee
where we have written CP$|s\rangle =|{\bar s}\rangle$ , with 
$|{\bar s}\rangle$ the antisoliton state, and 
CP$|\f\rangle =|\overline{\f}\rangle$.

Let us identify an appropriate final state.  First note that this is a
quantum process, so it is necessary to properly quantize the soliton
state\cite{ANW}.  This involves taking into account the zero modes of
the classical soliton configuration in order to construct a state
$|s\ra$ with definite quantum numbers.  The elements of the
translational zero mode may be superposed to create a definite
momentum state with center of mass momentum ${\bf P} = 0$.  Similarly,
the elements of the rotational/isorotational zero mode may be
superposed to create states with $I = J =
0,\frac{1}{2},1,\dots$.
Which choice of weak-isospin/spin is
appropriate depends on the specific physics underlying the Higgs
sector.  To be definite, here we choose $I = J = 0$.
If the soliton mass is not much larger than its inverse size, we
expect that the soliton will decay predominantly to states with a
small number of W-particles.  Consider the case where the final
state contains two $W_0$ particles (those particles that become
the $Z^0$'s when hypercharge is included), and where both particles have a
definite z-component of spin, $m = 1$.  Then, we define our final
state as
\be
	|\f\ra = |{\bf k},m_1=1;-{\bf k},m_2=1\ra.
\label{final}
\ee
Note that, with this choice, the CP conjugate state 
$|\overline{\f}\ra$ is identical to $|\f\ra$.

We can now identify a parameter which characterizes the CP violation.
Consider the difference between the differential probabilities to decay
into a specific two-particle final state
\bea
	\Delta_{CP} &\equiv& \left.\frac{d\Gamma}{d\Omega}(\cos 2\vp)
		\right|_{s\rightarrow|out\ra}
	      - \left.\frac{d\Gamma}{d\Omega}(\cos 2\vp)
		\right|_{\bar{s}\rightarrow|out\ra}		\nonumber
\eea
This may be expressed as

\bea
	\Delta_{CP} &=& \frac{k}{32\pi^2M^2_s}\left[|\la\f|{\cal T}|s\ra|^2
		- |\la\overline{\f}|{\cal T}|s\ra|^2\right]
								\nonumber \\
	&=& \frac{k\sin\theta}{16\pi^2M^2_s}\ 
		i({\cal A}^*_1{\cal A}_2 -{\cal A}^*_2{\cal A}_1) \ ,
\label{Delta1}
\eea
where $k = \sqrt{{\bf k}\cdot{\bf k}} = \sqrt{\frac{1}{4}{M_s}^2 - {m_W}^2}$
and $M_s$ is the mass of the soliton. Note that, in the limit in which there
is no CP violation ($\theta=0$), this quantity vanishes.

\section{Amplitudes from Semiclassical Methods}

\subsection{The Framework}
We perform the estimate of $\Delta_{CP}$ in the following way.  Describe
the quantum soliton as a coherent state around the classical soliton 
configuration, in the spherical ansatz\cite{sphanz}.
This coherent state tunnels along the minimal
Euclidean path underneath the appropriate energy barrier and, once the state 
emerges, it evolves along a path in configuration space corresponding to a
classical solution of the Euler-Lagrange equations.  Such a solution then
dissipates and asymptotically approaches a solution to the linearized 
spherical equations of motion.  We write these linearized classical solutions
as $f^n_{ia}({\bf k})$, where $n = 1,2$ labels the path. Thus, 
in unitary gauge, the field variables associated with
these classical solutions are

\be
A^n_{ia}(x)=\int d^3{\bf k}\, 
	[f^n_{ia}({\bf k})e^{-ik\cdot x} + f^{n*}_{ia}({\bf k})e^{ik\cdot x}]
\ .
\ee
Writing the coherent states built around $f^1_{ia}({\bf k})$ and 
$f^2_{ia}({\bf k})$ as $|f,1\ra$ and $|f,2\ra$ and taking the overlap 
between these states and the final state of interest, we obtain
\bea
	{\cal A}_1 &=& e^{iS_1}\la\f|f,1\ra	\nonumber \\
	{\cal A}_2 &=& e^{iS_2}\la\f|f,2\ra,
\label{conn}
\eea
where $e^{iS_1}$ and $e^{iS_2}$ are the respective saddle-point
evaluations of the coherent state path integral around the extremal
paths of type 1 and 2.  

The coherent state expresses $|f,1\ra$ and
$|f,2\ra$ as an expansion in Fock space with weights dependent on
$f^1_{ia}({\bf k})$ and $f^2_{ia}({\bf k})$, respectively.  After
some algebraic manipulation, the matrix elements $\la\f|f,n\ra$ can
be written
\be
\la{\bf k},m_1=1;-{\bf k},m_2=1|f,n\ra = B_n(k^2)\cos 2\vp \ ,
\label{overlap}
\ee
where $\vp$ is defined as the angle between the vector ${\bf k}$ and the
spin-quantization direction and

\be
	B_n(k^2) \equiv \frac{1}{2}\sqrt\frac{5}{32\pi}
			\left(3{\hat k}_i{\hat k}_j -
			\delta_{ij}\right) f^n_{ia}({\bf
			k})f^n_{ja}(-{\bf k}) \ .
\ee
Note that $B_n$ is only a function of $k^2$ since the $f^n_{ia}({\bf k})$ 
are in the spherical ansatz.  

Putting this all together yields

\be
	\Delta_{CP} = \frac{k\sin\theta}{16\pi^2M^2_s}(\cos 2\vp)^2\ 
	i\left[e^{i(S_2-S^*_1)}B^*_1B_2 -  e^{i(S_1-S^*_2)}B^*_2B_1\right]
\ .
\ee
It remains to estimate the quantities entering this expression in a 
controlled regime.

\subsection{The Limit}
In order to perform a controlled estimate, we work in the 
following limit:
$g \rightarrow 0$;
$m_W$ fixed;
$e$ fixed, implying $\xi = \frac{4e^2}{g^2} \rightarrow\infty$; and
$\lambda$ fixed, implying $m_H = \frac{\sqrt{8\lambda}}{g}\rightarrow\infty$.
Focus first on tunneling via path~1.
In the limit above, the gauge fields decouple from the soliton, and the 
properties of the soliton are completely determined by the Higgs dynamics.  
In particular, for the Skyrme model we are considering, this implies

\be
	M_s \sim \frac{v}{e},\ \  \ \ 
	L_s \sim \frac{1}{ev}.
\label{scales1}
\ee
The amplitude for Higgs unwinding remains unsupressed as $g \rightarrow 0$,
even though it is a tunneling process, and this implies that 
${\rm Re}(S_1) \sim {\cal O}(1)$ and ${\rm Im}(S_1) \sim {\cal O}(1)$ in the
expression~(\ref{conn}).  Moreover, because the mass and length of the soliton
scale as in (\ref{scales1}), there is no further supression from
$\la\f|f,1\ra$ in (\ref{overlap}).

Now focus on  tunneling via path~2. Consider the dynamics in the unitary gauge.
By rescaling the action, we see that $g^2$ plays a role analogous to 
$\hbar$, and that the soliton approaches 
a pure winding configuration of size $[m_W\sqrt{\xi}]^{-1}$.  
Similarly, the Euclidean tunneling path approaches the 't Hooft 
instanton\cite{tHooft} of the same size, implying that the decay amplitude 
approaches the instanton amplitude\cite{RST 85}.  Because the configuration 
that emerges from under
the barrier at the end of tunneling is governed by the same
length scale, $[m_W\sqrt{\xi}]^{-1}$, the number of quanta in the final state
is approximately

\be
	N_{quanta} \sim \frac{E}{L^{-1}} \sim 
		\frac{m_W/(g^2\sqrt{\xi})}{m_W\sqrt{\xi}}  \sim {\cal O}(1)
\ .
\ee
Thus, the amplitude is not further supressed by having a small number of
particles in the final state, implying $\la\f|f,2\ra \sim {\cal O}(1)$.
Moreover, the configuration that emerges from under the barrier is already
linearized, which implies that in (\ref{conn}),
$e^{iS_2}\sim {\cal O}(e^{-8\pi^2/g^2})$ is real.

Thus, our estimate yields
\be
	|\Delta(\cos 2\vp)| \sim e^{-8\pi^2/g^2}(\cos 2\vp)^2\sin\theta \ .
\label{leading}
\ee
for the difference between the differential decay probability for soliton
decay to a two $W_0$-particle $S=2,M_S=2$ state and that for antisoliton
decay to the same two $W_0$-particle state.

\section{Concluding Remarks}

In this paper, we have presented a relatively simple scenario by which a total
divergence manifests itself in a physically observable processes.  We
considered the effect of adding a CP-odd total divergence to the electoweak
theory without fermions.  We showed that if Higgs winding solitons exist in
such a theory, their decay into a specific final state violates CP 
through the interference of two topologically distinct decay channels.
Finally, for consistency, we should demonstrate how the effect disappears when 
fermions are included.

If one introduces fermions, the axial anomaly implies that
\be
\partial_{\mu}J_F^{\mu} = F\tilde{F} \ ,
\ee
so that in any process for which $\int F\tilde{F} \neq 0$, fermions are
created or destroyed. this implies that in the Hamiltonian picture the 
degeneracy of vacua with different winding is lifted and thus any 
interference between paths 1 and 2 goes away. Thus, any CP violation
ceases to exist.

One may also describe the effect of the $\theta F\tilde{F}$-term as
the inclusion of an extra phase on the fermion state, since the phase
only appears when a fermion is produced.  Thus, the role of the
$\theta$-term may be considered as a redefinition of the phase of
the fermion state.  It should be no suprise that this is precisely
the mechanics by which one formally removes the term from the functional
integral\cite{Fujikawa}.

\acknowledgments
We would like to thank to Eddie Farhi for collaborating with us on many of
the ideas presented in this paper, and for providing valuable insights into 
the effects of total divergences in field theories.
We also wish to thank Jeffrey Goldstone, Ken Johnson, Al Mueller, Lisa 
Randall, and Tanmay Vachaspati for helpful discussions.
A.L. was supported by funds provided by the US Department of Energy (D.O.E.)
and M.T. was supported by the D.O.E, the National Science foundation, 
and by funds provided by Case Western Reserve University.

\end{document}